\documentclass[twocolumn]{aastex62}
\shorttitle{SRBs from Galactic magnetars?}
\shortauthors{Zhang}
\begin{document}


\title{\bf ``Slow'' Radio Bursts from Galactic Magnetars?}

%% Use \author, \affil, and the \and command to format author and affiliation 
%% information.  If done correctly the peer review system will be able to
%% automatically put the author and affiliation information from the manuscript
%% and save the corresponding author the trouble of entering it by hand.
%%
%% The \affil should be used to document primary affiliations and the
%% \altaffil should be used for secondary affiliations, titles, or email.

%% Authors with the same affiliation can be grouped in a single
\author{Bing Zhang }
\affil{Department of Physics and Astronomy, University of Nevada, Las Vegas, Las Vegas, NV 89154,
zhang@physics.unlv.edu}

\begin{abstract}
Recently, one fast radio burst, FRB 200428, was detected from the Galactic magnetar SGR J1935+2154 during one X-ray burst. This suggests that magnetars can make FRBs. On the other hand, the majority of X-ray bursts from SGR J1935+2154 are not associated with FRBs. One possible reason for such rarity of FRB-SGR-burst associations is that the FRB emission is much more narrowly beamed than the SGR burst emission. If such an interpretation is correct, one would expect to detect radio bursts with viewing angles somewhat outside the narrow emission beam. These ``slow'' radio bursts (SRBs) would have  broader widths and lower flux densities due to the smaller Doppler factor involved. We derive two ``closure relations'' to judge whether a long, less luminous radio burst could be an SRB. The $2.2$-s, $308 \ {\rm Jy \ ms}$, $\sim 111$ MHz radio burst detected from SGR J1935+2154 by the BSA LPI radio telescope may be such an SRB. The $\sim 2$-ms, $60 \ {\rm mJy \ ms}$ faint burst detected by FAST from the same source could be also an SRB if the corresponding FRB has a narrow spectrum.  If the FRB beam is narrow, there should be many more SRBs than FRBs from Galactic magnetars. The lack of detection of abundant SRBs from magnetars would disfavor the hypothesis that all SGR bursts are associated with narrow-beam FRBs. 
 \end{abstract}

\keywords{radio transient sources -- magnetars}

\section{Introduction} \label{sec:intro}

The detection of a $1.5 \ {\rm MJy \ ms}$ fast radio burst (FRB) in the Milky Way galaxy, i.e. FRB 200428 \citep{STARE2-SGR,CHIME-SGR} in association with a bright X-ray burst \citep{HXMT-SGR,Integral-SGR,AGILE-SGR,Konus-SGR} from the magnetar SGR J1935+2154, established the magnetar origin of at least some, probably all FRBs \citep{popov10,lyubarsky14,kulkarni14,katz16,metzger17,beloborodov17,kumar17,yangzhang18,wadiasingh20,lu20,margalit20,yang20c,zhang20b}. 
On the other hand, deep monitoring of SGR J1935+2154 by Five-hundred-meter Aperture Spherical radio Telescope (FAST) \citep{lin20} suggested that the FRB-SGR-burst associations are rather rare. During an active phase of SGR J1935+2154 when 29 other X-ray bursts were emitted from the source, no single FRB-like event was detected. Whereas whether the FRB-associated X-ray burst is physically special is still subject to debate \citep[e.g.][]{HXMT-SGR,NICER-SGR,yangyh20}, one plausible possibility is that the FRB emission is much more narrowly beamed than the SGR burst emission \citep{lin20}. 

Within this picture, an FRB can be detected only when the narrow beam points towards Earth. Outside the FRB ``jet'', due to the rapid drop of the Doppler factor, one would expect that the flux drops rapidly, spectrum becomes softer, and duration becomes longer. These off-beam events are not likely detectable from cosmological FRB sources. However,  in view of the huge specific fluence of FRB 200428, it is entirely possible that some off-axis, longer and softer bursts from Galactic magnetars such as SGR J1935+2154 can be detected above the sensitivity thresholds of the available radio telescopes. We define these events as ``slow'' radio bursts (SRBs) and study their properties in this {\em Letter}. 

\section{On-beam FRB vs. off-beam SRB}

FRB emission models from magnetars invoke either magnetospheres \citep{kumar17,yangzhang18,wadiasingh20,lu20,yang20c} or relativistic shocks \citep{lyubarsky14,metzger17,metzger19,beloborodov17,beloborodov20,margalit20} to produce FRBs. Both types of models invoke a relativistically moving plasma to produce FRB emission \citep[][and references therein]{zhang20b}. Regardless of the emission site, here we consider a relativistically moving conical jet with bulk Lorentz factor $\Gamma$ (and dimensionless speed $\beta$) and half opening angle $\theta_j$. Consider an observer at a viewing angle $\theta$ from the jet axis, in general one can define the Doppler factor
\begin{equation}
 {\cal D (\theta)} = \left\{ 
   \begin{array}{ll}
   {\cal D}_{\rm on} = \frac{1}{\Gamma(1-\beta)} \simeq 2\Gamma, & \theta \leq \theta_j, \nonumber \\
   {\cal D}_{\rm off} =  \frac{1}{\Gamma(1-\beta \cos(\Delta\theta))}, & \theta > \theta_j,
   \end{array}
 \right. 
 \label{eq:Doppler}
\end{equation}
where $\Delta \theta = \theta - \theta_j$. 
The Doppler factor makes a connection between co-moving-frame quantities (primed) and the observer-frame quantities \citep[e.g.][]{zhang18}
\begin{eqnarray}
 \nu & = & {\cal D} \nu',  \nonumber \\
 \Delta t & = & {\cal D}^{-1} \Delta t', \nonumber \\ 
 L_\nu  & = & {\cal D}^3 L'_{\nu'},
\end{eqnarray}
where $\nu$ is the emission frequency, $\Delta t$ is the emission duration, and $L_\nu$ is the isotropic-equivalent specific luminosity. The last equation makes use of the point source assumption, which is justified for FRB sources.

We compare the observed properties of two observers, one on-beam observer with $\theta \leq \theta_j$ and ${\cal D}={\cal D}_{\rm on}$ and another off-beam observer with $\theta>\theta_j$ and ${\cal D}={\cal D}_{\rm off}$. Define a Doppler factor ratio
\begin{equation}
 {\cal R}_{\cal D} \equiv \frac{{\cal D}_{\rm on}}{{\cal D}_{\rm off}} > 1
 \label{eq:RD}
\end{equation}
and assume that the comoving-frame parameters are the same for on-beam and off-beam observers, one can write down the relationships of the properties of an on-beam FRB and an off-beam SRB.

Let us consider a radio burst with comoving frame full-width at half maximum (FWHM) of $w'$ and specific luminosity power-law spectrum 
\begin{equation}
L'_{\nu'} (\nu')= L'_{\nu'} (\nu'_0) \left(\frac{\nu'}{\nu'_0}\right)^{-\alpha}
\label{eq:spectrum}
\end{equation}
at the burst peak time, where $\nu'_0$ is a characteristic frequency and $\alpha$ is the spectral index. Consider an off-axis observer observing at $\nu_2$ and an on-axis observer observing at $\nu_1$, the ratio between the specific luminosities of the two observers reads
\begin{eqnarray}
 \frac{L_\nu^{\rm off}(\nu_2)}{L_\nu^{\rm on}(\nu_1)} & = & \frac{L_\nu(\nu_0^{\rm off})}{L_\nu(\nu_0^{\rm on})} \left(\frac{\nu_0^{\rm on}}{\nu_0^{\rm off}}\right)^{-\alpha} \left( \frac{\nu_2}{\nu_1} \right)^{-\alpha} \nonumber \\
 & = & {\cal R}_{\cal D}^{-3-\alpha}  \left( \frac{\nu_2}{\nu_1} \right)^{-\alpha}.
\end{eqnarray}
Noticing 
\begin{equation}
\frac{w^{\rm off}}{w^{\rm on}} = {\cal R}_{\cal D}
\label{eq:w-ratio}
\end{equation} 
and considering that the specific luminosity ratio is proportional to the specific flux ratio, we finally get a ``closure relation'' among the ratios of specific fluence ${\cal F}_\nu$, width $w$, and the observing frequency $\nu$ between an off-axis SRB and an on-axis FRB:
\begin{equation}
 \left( \frac{{\cal F}_\nu^{\rm SRB}}{{\cal F}_\nu^{\rm FRB}} \right) \left(\frac{w^{\rm SRB}}{w^{\rm FRB}}\right)^{2+\alpha} \left(\frac{\nu^{\rm SRB}}{\nu^{\rm FRB}}\right)^{\alpha} = 1,
\label{eq:closure}
\end{equation}
where ${\cal F}_\nu^{\rm SRB} = L_\nu^{\rm off}(\nu_2) w^{\rm off} / 4\pi d_{\rm L}^2$, ${\cal F}_\nu^{\rm FRB} = L_\nu^{\rm on}(\nu_1) w^{\rm on} / 4\pi d_{\rm L}^2$, $w^{\rm SRB} = w^{\rm off}$, $w^{\rm FRB} = w^{\rm on}$, $\nu^{\rm SRB}=\nu_2$, and $\nu^{\rm FRB}=\nu_1$. This relation can be used to determine whether a long-duration, low-fluence burst could be the off-beam version of an FRB.

The closure relation (\ref{eq:closure}) makes the assumption of a power law spectrum for FRBs and depends on the spectral index $\alpha$. The true FRB spectral shape is not well measured. In any case, in the narrow band where FRBs are detected, some FRBs show a power law spectrum, but with a wide range of spectral index, e.g. $\alpha = 4 \pm 1$ for the Lorimer burst \citep{lorimer07}, $\alpha = 7.8 \pm 0.4$ for FRB 110523 \citep{masui15}, $\alpha = -0.3 \pm 0.9$ for FRB 131104 \citep{ravi15}, and $\alpha$ ranging from -10.4 to +13.6 for FRB 121102 \citep{spitler16}. 
 
There is evidence that at least some FRB spectra may be narrow \citep{spitler16}. The closure relation  (\ref{eq:closure}) also relies on the assumption that the FRB spectra extend to a high-enough frequency in the rest-frame, so that the same index still applies for off-beam events. If the FRB spectrum is curved and intrinsically narrow, the closure relation could be more complicated. As an example, we consider a rest-frame  Gaussian-like spectrum, i.e.
\begin{equation}
 L'_{\nu'}(\nu') = L'_{\nu'} (\nu'_0) \exp \left[-\frac{1}{2} \left( \frac{\nu'-\nu'_0}{\delta \nu'} \right)^2 \right],
 \label{eq:spectrum-2}
\end{equation}
where $\nu'_0$ and $\delta\nu'$ are the center frequency and the characteristic width of the spectrum, respectively. For a narrow spectrum, one has $\delta \nu' \ll \nu'_0$. For an FRB with specific fluence ${\cal F}_\nu^{\rm FRB}$, intrinsic width $w^{\rm FRB}$ and observing frequency at $\nu_1$, one may assume $\nu_1 = {\cal D}_{\rm on} \nu'_0 = \nu^{\rm FRB}$ and $\delta \nu^{\rm on} = {\cal D}_{\rm on} \delta \nu' = \delta\nu^{\rm FRB}$. An SRB observed at a large viewing angle would satisfy a more complicated closure relation (see Appendix)
\begin{eqnarray}
 & \left( \frac{{\cal F}_\nu^{\rm SRB}}{{\cal F}_\nu^{\rm FRB}} \right)  \left(\frac{w^{\rm SRB}}{w^{\rm FRB}}\right)^{2} \nonumber \\
&  \times  \exp \left[\frac{1}{2} \left(\frac{\nu^{\rm FRB}}{\delta \nu^{\rm FRB}} \right)^2 \left(\frac{w^{\rm SRB} \nu^{\rm SRB}}{w^{\rm FRB} \nu^{\rm FRB}} - 1 \right)^2 \right] = 1.
\label{eq:closure-2}
\end{eqnarray}
If $\delta\nu^{\rm FRB} \ll \nu^{\rm FRB}$, specific fluence drops significantly if the frequency is beyond the width of the line spectrum at an off-beam angle. Rapid variability is expected for such narrow-spectra FRBs \citep{beniamini20}.
 
\section{SRB properties and case studies}

From Equations (\ref{eq:closure}) and (\ref{eq:closure-2}), one may predict the properties of an SRB based on known properties of an FRB. A typical FRB with a specific fluence ${\rm 1 \ Jy \ ms}$ at a 100 Mpc cosmological distance would have a specific fluence of ${\rm 100 \ MJy \ ms}$ at a typical Galactic distance of 10 kpc. Assuming the same telescope ($\nu^{\rm SRB}=\nu^{\rm FRB}$) and for a power law spectrum, one gets the SRB width longer than the FRB width by a factor of $10^{8/(2+\alpha)}$ if it also has a specific fluence of ${\rm 1 \ Jy \ ms}$ level. This would make the burst duration of the order of a second for $\alpha=1$ and even 10 seconds for a flat spectrum burst. So the main characteristic of an SRB is its ``slow'' nature.  By definition in Eq.(\ref{eq:RD}), the discussion of an SRB is only relevant when its width is longer than its corresponding FRB. 

Below we discuss several interesting radio bursts detected from SGR J1935+2154 and their compliance with the SRB closure relations.

\subsection{The BSA LPI burst}
The BSA/LPI radio telescope at Pushchino Radio Astronomy Observatory, Russia, detected one radio burst from SGR J1935+2154 at 2020-09-02 UTC 18:14:59 at 111 MHz with a 2.5 MHz band \citep{rodin20}. The measured dispersion measure DM is in general consistent with the DM measured from other radio telescopes \citep{CHIME-SGR,STARE2-SGR,zhangc20,zhu20}. The burst has a measured pulse width of 2.2 s, a flux density of 140 mJy, and a specific fluence $308 \ {\rm Jy \ ms}$ with an observing frequency of 111 MHz. The intrinsic width is $\sim 340$ ms after correcting for the scattering and instrumental effects (A. Rodin, 2020, private communication). We test the SRB hypothesis by setting ${\cal F}_\nu^{\rm SRB} \simeq 308 \ {\rm Jy \ ms}$, $w^{\rm SRB}=0.34$ s, and $\nu^{\rm SRB} = 111$ MHz. We test the closure relations against two reference FRBs: (1) Reference FRB-1 as a typical cosmological FRB with ${\cal F}_\nu^{\rm FRB} \simeq 100 \ {\rm MJy \ ms}$, $w^{\rm FRB} \simeq 1$ ms, and $\nu^{\rm FRB} \simeq 1.2$ GHz if it were detected in the Milky Way; and (2) Reference FRB-2 as FRB 200428 itself, with ${\cal F}_\nu^{\rm FRB} \simeq 1.5\ {\rm MJy \ ms}$ and $w^{\rm FRB} \simeq 0.61$ ms at $\nu^{\rm FRB} \simeq 1.52$ GHz \citep{STARE2-SGR}. 

The Gaussian-like spectrum closure relation, Equation (\ref{eq:closure-2}), can be made to match Reference FRB-1, given that $\delta\nu^{\rm FRB} / \nu^{\rm FRB} \sim 21$, suggesting that the spectrum is quite wide. There is no solution for Reference FRB-2, suggesting that this burst cannot be the corresponding SRB of FRB 200428 if the spectrum of FRB 200428 is Gaussian-like. For the power-law spectrum closure relation Eq. (\ref{eq:closure}),  Reference FRB-1 requires $\alpha=0.30$ while Reference FRB-2 requires $\alpha=-1.1$. These are reasonable $\alpha$ values for known FRBs \citep{lorimer07,masui15,ravi15,spitler16}. We therefore conclude that the long duration radio burst detected by the BSA/LPI radio telescope \citep{rodin20} is consistent with being an SRB.

\subsection{Other weak bursts from SGR J1935+2154}

Shortly after the detection of FRB 200428, \cite{zhangc20} reported the detection of a weak burst from SGR J1935+2154 with FAST with central frequency 1.25 GHz. The relevant measurements are: specific fluence ${\cal F}_\nu \sim 60 \ {\rm mJy \ ms} = 0.06 {\rm Jy \ ms}$, $w \sim 1.97$ ms, and $\nu \sim 1.25$ GHz. \cite{yu20} speculated that it could be an off-beam FRB within the framework of the synchrotron maser shock model. We confront the data with the SRB closure relations. For the case of a power law spectrum closure relation (\ref{eq:closure}), we find that very extreme spectral indices are needed: $\alpha \sim 27.7$ for Reference FRB-1 and $\alpha \sim 15.0$ for Reference FRB-2. We then check the Gaussian-like spectrum closure relation (\ref{eq:closure-2}). It can be satisfied for Reference FRB-1 if $\delta\nu^{\rm FRB} / \nu^{\rm FRB} \sim 0.17$ and for Reference FRB-2 if $\delta\nu^{\rm FRB} / \nu^{\rm FRB} \sim 0.44$. As a result, this burst could be an SRB only if the FRB spectrum is not a power law, but a narrow Gaussian. There was no X-ray burst associated with this burst. If the SRB interpretation is correct, then either this SRB is not associated with an X-ray burst, or the X-ray burst is not much broader than the corresponding FRB itself.

\cite{kirsten20} reported two more bursts (Wb B1 and B2) from SGR J1935+2154 detected with the Westerbork RT1 radio telescope with central frequency $\nu = 1.324$ GHz. The relevant measurements are: ${\cal F}_\nu \sim 112 \ {\rm Jy \ ms}$ and $w \sim 0.427$ ms for B1, and ${\cal F}_\nu \sim 24 \ {\rm Jy \ ms}$ and $w \sim 0.219$ ms for B2, respectively. Since both bursts have widths shorter than both reference FRBs, we immediately draw the conclusion that they cannot be the corresponding SRBs of the two reference FRBs. The possibility that they are the SRB counterparts of an intrinsically narrower FRB is still possible. For example, if we consider a new, Reference FRB-3 as a cosmological FRB with ${\cal F}_\nu^{\rm FRB} \simeq 100 \ {\rm MJy \ ms}$, $w^{\rm FRB} \simeq 0.1$ ms, and $\nu^{\rm FRB} \simeq 1.2$ GHz, solutions can be found to satisfy both closure relations for both bursts. For Wb B1, one requies $\alpha=6.97$ for the power law model and  $\delta\nu^{\rm FRB} / \nu^{\rm FRB} \sim 0.8$ for the Gaussian model; for Wb B2, one requires  $\alpha=15.5$ for the power law model and $\delta\nu^{\rm FRB} / \nu^{\rm FRB} \sim 0.27$ for the Gaussian model. 

Even if the possibility that these weak bursts from SGR J1935+2154 are SRBs is not ruled out, it is more likely that they are intrinsically weaker bursts emitted by SGR J1935+2154. If so, magnetars can generate bursting radio emission in a very wide luminosity range. This poses interesting constraints on the coherent mechanisms for magnetar radio emission.

\section{Detectability}

In order to detect an SRB with specific fluence above the telescope's sensitivity threshold, the viewing angle should not be too far outside the FRB jet cone. The maximum viewing angle $\theta_{\rm max}$ depends on the typical $\Gamma$ and $\theta_j$ of the FRB emitter as well as the shape of the FRB spectrum. For simplicity, we consider the same telescope to detect FRBs and SRBs so that $\nu^{\rm SRB}=\nu^{\rm FRB}$. The maximum viewing angle defines $\Delta \theta_{\rm max} = \theta_{\rm max}-\theta_j$, at which ${\cal F}_\nu^{\rm SRB}$ equals the fluence sensitivity threshold ${\cal F}_{\nu,th}$ of the telescope. For simplicity, we consider a power law spectrum. Making use of Equations (\ref{eq:w-ratio}) and (\ref{eq:Doppler}), Equation (\ref{eq:closure}) can be then re-written as 
\begin{equation}
 \beta\cos(\Delta\theta_{\rm max})=1-\frac{1}{2\Gamma^2} \left(\frac{{\cal F}_{\nu}^{\rm FRB}}{{\cal F}_{\nu,th}} \right)^{1/(2+\alpha)}.
\end{equation}
For $\Gamma \gg 1$ ($\beta \simeq (1-1/2\Gamma^2)$) and $\theta_{\rm max} \ll 1$, this can be reduced to
\begin{eqnarray}
 \Delta \theta_{\rm max} & \simeq & \frac{1}{\Gamma} \left[ \left( \frac{{\cal F}_{\nu}^{\rm FRB}} { {\cal F}_{\nu,th}}  \right)^{1/(2+\alpha)} -1  \right]^{1/2} \nonumber \\
 & \simeq &  \frac{1}{\Gamma} \left( \frac{{\cal F}_{\nu}^{\rm FRB}} { {\cal F}_{\nu,th}}  \right)^{1/(4+2\alpha)} =  \frac{\xi}{\Gamma} \gg \frac{1}{\Gamma},
\end{eqnarray}
where $\xi= \left( {{\cal F}_{\nu}^{\rm FRB}}/ { {\cal F}_{\nu,th}}  \right)^{1/(4+2\alpha)}  \gg 1$. For ${\cal F}_{\nu}^{\rm FRB} / {\cal F}_{\nu,th} \simeq 10^8$, one has $\xi \simeq 100$ for $\alpha=0$ and $\xi \simeq 10$ for $\alpha=2$.  The small angle approximation is valid only when $\Gamma \gg 100$ or $\Gamma \gg 10$ in the respective cases.  

The solid angle ratio between detectable Galactic SRBs and FRBs can be estimated as
\begin{eqnarray}
 {\cal R}_{\Delta\Omega} & \equiv & \frac{\Delta\Omega^{\rm SRB}}{\Delta\Omega^{\rm FRB}} \simeq \frac{\pi [(\theta_j+\Delta\theta_{\rm max})^2 - \theta_j^2]}{\pi \theta_j^2} \nonumber \\
 & = & \left( \frac{\Delta\theta_{\rm max}}{\theta_j} \right)^2 + 2 \left( \frac{\Delta\theta_{\rm max}}{\theta_j} \right) \nonumber \\
 & = & \left( \frac{\xi}{\Gamma\theta_j} \right)^2 + \left(\frac{2\xi}{\Gamma\theta_j}\right). 
\end{eqnarray}
This is also the ratio of event rates of the detectable SRBs and FRBs from Galactic magnetars. One can see that for wide-beam FRBs with $\theta_j \gg \xi/\Gamma$, which is relevant for relativistic shock models, the Galactic SRB rate would not be higher than the Galactic FRB rate. However, for narrow beams $\theta_j \ll \xi/\Gamma$ which is more relevant for a magnetospheric origin of FRBs, there should be many more SRBs than FRBs if the beaming interpretation of the FRB paucity is valid. If detailed data analyses suggest that SRBs are not as abundant as FRBs from Galactic magnetars, it may suggest either of the following two scenarios. First, the FRB beams may be very wide, which could be consistent with the models invoking emission beyond the magnetar light cylinder. Such models, on the other hand, face the challenge of interpreting diverse polarization angle variations from some repeating FRBs \citep{luo20,zhang20b}. Alternatively, if FRB emission indeed originates from narrow beams within magnetar magnetospheres, the lack of detection of abundant SRBs would suggest that the rarity of FRB-SGR-burst associations is likely intrinsic, i.e., the SGR-burst that makes FRB 200428 was physically distinct from other SGR bursts \citep{HXMT-SGR,NICER-SGR,yangyh20}. 

\section{Conclusions and discussion}

We have discussed a type of radio burst from Galactic magnetars that could be FRBs viewed off-beam. These bursts, dubbed SRBs, could have much longer durations and lower specific fluences than FRBs because of their smaller Doppler factors than on-beam FRBs. We derive two ``closure relations'', Equations (\ref{eq:closure}) and (\ref{eq:closure-2}), among the ratios of burst specific fluence, width, and observing frequency between SRBs and FRBs,  which could be used to judge whether a radio burst is an SRB. We show that The 2.2-s long, 111 MHz radio burst detected from SGR J1935+2154 by the BSA LPI radio telescope \citep{rodin20} could be interpreted as an SRB. The weak FAST burst \citep{zhangc20} may be also interpreted as an SRB if the corresponding FRB has a narrow, Gaussian-like spectral shape.

We estimate the relative event rates of Galactic SRBs with respect to Galactic FRBs. The rate of Galactic SRBs could be much higher than that of Galactic FRBs if all SGR-bursts are associated with narrow-beam FRBs \citep{lin20}. A systematic search for SRBs from SGR J1935+2154 and other Galactic magnetars can place important constraints on this hypothesis. The lack of detection of abundant Galactic SRBs would rule out the hypothesis that all SGR X-ray bursts are associated with narrow-beam FRBs. Identifications of SRBs from Galactic magnetars, on the other hand, would confirm the beaming nature of FRBs and allow direct constraints on the physical parameters of the FRB emitters. 

SRBs may not be only produced by Galactic magnetars. If other sources in the Milky Way can make Galactic FRBs and be detected by future wide-field radio telescope arrays, SRBs may be also generated from those objects based on the same reasoning discussed here. 

\acknowledgments 
The author thanks Alexander Rodin for clarifying the observational details of the BSA/LPI radio burst, Yuan-Pei Yang and Yun-Wei Yu for comments, and an anonymous referee for helpful suggestions.

\appendix

\section{Derivation of Equation (9)}

With a Gaussian-like spectrum in Eq.(\ref{eq:spectrum-2}), one can derive
\begin{equation}
  \frac{L_\nu^{\rm off}(\nu_2)}{L_\nu^{\rm on}(\nu_1)}  =  \frac{{\cal D}_{\rm off}^3}{{\cal D}_{\rm on}^3} \frac{\exp \left[ -\frac{1}{2} \left( \frac{\nu_2 - \nu_0^{\rm off}}{\delta \nu^{\rm off}} \right)^2 \right]}{\exp \left[ -\frac{1}{2} \left( \frac{\nu_2 - \nu_0^{\rm on}}{\delta \nu^{\rm on}} \right)^2 \right]}  
  = {\cal R}_{\cal D}^{-3}  \exp \left\{ - \frac{1}{2} \left[ \left(\frac{\nu'_2 - \nu'_0}{\delta\nu'}\right)^2 - \left(\frac{\nu'_1 - \nu'_0}{\delta\nu'}\right)^2 \right] \right\}.
  \label{eq:A1}
\end{equation}
For an FRB, one can assume that the observing frequency is right at the center frequency of the spectrum, i.e. $\nu'_1 = \nu'_0$. Equation (\ref{eq:A1}) now becomes 
\begin{equation}
  \frac{L_\nu^{\rm off}(\nu_2)}{L_\nu^{\rm on}(\nu_1)} 
  = {\cal R}_{\cal D}^{-3}  \exp \left[ - \frac{1}{2} \left(\frac{\nu'_2 - \nu'_0}{\delta\nu'}\right)^2 \right]  = {\cal R}_{\cal D}^{-3}  \exp \left[ - \frac{1}{2} \left(\frac{ \frac{\nu_2}{{\cal D}_{\rm off}} - \frac{\nu_1}{{\cal D}_{\rm on}}}{\frac{\delta\nu^{\rm on}}{{\cal D}_{\rm on}}}\right)^2 \right] = {\cal R}_{\cal D}^{-3}  \exp\left[ -\frac{1}{2}  \frac{\nu_1^2}{(\delta\nu^{\rm on})^2} \left({\cal R}_{\cal D}\frac{\nu_2}{\nu_1} -1 \right)^2 \right].
  \label{eq:A2}
\end{equation}
Following the same logic of deriving Eq.(\ref{eq:closure}), one can then derive Eq.(\ref{eq:closure-2}).

%\bibliography{FRB}

\end{document}